\documentclass[twoside]{article}
\input{epsf}

\newcommand{\eg}{{\em e.g.\ }}
\newcommand{\vs}{{\em vs.\ }}
\catcode`\@=11
\long\def\@makefntext#1{
\protect\noindent \hbox to 3.2pt {\hskip-.9pt  
$^{{\eightrm\@thefnmark}}$\hfil}#1\hfill}		

\def\@makefnmark{\hbox to 0pt{$^{\@thefnmark}$\hss}}	
	
\def\ps@myheadings{\let\@mkboth\@gobbletwo
\def\@oddhead{\hbox{}
\rightmark\hfil\eightrm\thepage}   
\def\@oddfoot{}\def\@evenhead{\eightrm\thepage\hfil
\leftmark\hbox{}}\def\@evenfoot{}
\def\sectionmark##1{}\def\subsectionmark##1{}}

\oddsidemargin=\evensidemargin
\newcounter{sectionc}\newcounter{subsectionc}\newcounter{subsubsectionc}
\renewcommand{\section}[1] {\vspace{12pt}\addtocounter{sectionc}{1} 
\setcounter{subsectionc}{0}\setcounter{subsubsectionc}{0}\noindent 
	{\tenbf\thesectionc. #1}\par\vspace{5pt}}
\renewcommand{\subsection}[1] {\vspace{12pt}\addtocounter{subsectionc}{1} 
	\setcounter{subsubsectionc}{0}\noindent 
	{\bf\thesectionc.\thesubsectionc. {\kern1pt \bfit #1}}\par\vspace{5pt}}
\renewcommand{\subsubsection}[1] {\vspace{12pt}\addtocounter{subsubsectionc}{1}
	\noindent{\tenrm\thesectionc.\thesubsectionc.\thesubsubsectionc.
	{\kern1pt \tenit #1}}\par\vspace{5pt}}

\newcounter{appendixc}
\newcounter{subappendixc}[appendixc]
\newcounter{subsubappendixc}[subappendixc]
\renewcommand{\thesubappendixc}{\Alph{appendixc}.\arabic{subappendixc}}
\renewcommand{\thesubsubappendixc}
	{\Alph{appendixc}.\arabic{subappendixc}.\arabic{subsubappendixc}}

\renewcommand{\appendix}[1] {\vspace{12pt}
        \refstepcounter{appendixc}
        \setcounter{figure}{0}
        \setcounter{table}{0}
        \setcounter{lemma}{0}
        \setcounter{theorem}{0}
        \setcounter{corollary}{0}
        \setcounter{definition}{0}
        \setcounter{equation}{0}
        \renewcommand{\thefigure}{\Alph{appendixc}.\arabic{figure}}
        \renewcommand{\thetable}{\Alph{appendixc}.\arabic{table}}
        \renewcommand{\theappendixc}{\Alph{appendixc}}
        \renewcommand{\thelemma}{\Alph{appendixc}.\arabic{lemma}}
        \renewcommand{\thetheorem}{\Alph{appendixc}.\arabic{theorem}}
        \renewcommand{\thedefinition}{\Alph{appendixc}.\arabic{definition}}
        \renewcommand{\thecorollary}{\Alph{appendixc}.\arabic{corollary}}
        \renewcommand{\theequation}{\Alph{appendixc}.\arabic{equation}}
        \noindent{\tenbf Appendix \theappendixc #1}\par\vspace{5pt}}
\newcommand{\subappendix}[1] {\vspace{12pt}
        \refstepcounter{subappendixc}
        \noindent{\bf Appendix \thesubappendixc. {\kern1pt \bfit #1}}
	\par\vspace{5pt}}
\newcommand{\subsubappendix}[1] {\vspace{12pt}
        \refstepcounter{subsubappendixc}
        \noindent{\rm Appendix \thesubsubappendixc. {\kern1pt \tenit #1}}
	\par\vspace{5pt}}

\newcommand{\smalllineskip}{\baselineskip=10pt}

\def\eightcirc{
\begin{picture}(0,0)
\put(4.4,1.8){\circle{6.5}}
\end{picture}}
\def\eightcopyright{\eightcirc\kern2.7pt\hbox{\eightrm c}}

\renewenvironment{thebibliography}[1]
	{\frenchspacing
	 \ninerm\baselineskip=11pt
	 \begin{list}{\arabic{enumi}.}
	{\usecounter{enumi}\setlength{\parsep}{0pt}
	 \setlength{\leftmargin 12.7pt}{\rightmargin 0pt} 
	 \setlength{\itemsep}{0pt} \settowidth
	{\labelwidth}{#1.}\sloppy}}{\end{list}}

\newcounter{itemlistc}
\newcounter{romanlistc}
\newcounter{alphlistc}
\newcounter{arabiclistc}

\newcommand{\fcaption}[1]{
        \refstepcounter{figure}
        \setbox\@tempboxa = \hbox{\footnotesize Fig.~\thefigure. #1}
        \ifdim \wd\@tempboxa > 5in
           {\begin{center}
        \parbox{5in}{\footnotesize\smalllineskip Fig.~\thefigure. #1}
            \end{center}}
        \else
             {\begin{center}
             {\footnotesize Fig.~\thefigure. #1}
              \end{center}}
        \fi}

\newcommand{\tcaption}[1]{
        \refstepcounter{table}
        \setbox\@tempboxa = \hbox{\footnotesize Table~\thetable. #1}
        \ifdim \wd\@tempboxa > 5in
           {\begin{center}
        \parbox{5in}{\footnotesize\smalllineskip Table~\thetable. #1}
            \end{center}}
        \else
             {\begin{center}
             {\footnotesize Table~\thetable. #1}
              \end{center}}
        \fi}

\def\@citex[#1]#2{\if@filesw\immediate\write\@auxout
	{\string\citation{#2}}\fi
\def\@citea{}\@cite{\@for\@citeb:=#2\do
	{\@citea\def\@citea{,}\@ifundefined
	{b@\@citeb}{{\bf ?}\@warning
	{Citation `\@citeb' on page \thepage \space undefined}}
	{\csname b@\@citeb\endcsname}}}{#1}}

\newif\if@cghi
\def\cite{\@cghitrue\@ifnextchar [{\@tempswatrue
	\@citex}{\@tempswafalse\@citex[]}}
\def\citelow{\@cghifalse\@ifnextchar [{\@tempswatrue
	\@citex}{\@tempswafalse\@citex[]}}
\def\@cite#1#2{{$\null^{#1}$\if@tempswa\typeout
	{IJCGA warning: optional citation argument 
	ignored: `#2'} \fi}}

\def\pmb#1{\setbox0=\hbox{#1}
	\kern-.025em\copy0\kern-\wd0
	\kern.05em\copy0\kern-\wd0
	\kern-.025em\raise.0433em\box0}

\def\fnt#1#2{\footnotetext{\kern-.3em
	{$^{\mbox{\scriptsize #1}}$}{#2}}}

\font\tenrm=cmr10
\font\tenit=cmti10 
\font\tenbf=cmbx10
\font\bfit=cmbxti10 at 10pt
\font\ninerm=cmr9

\font\eightrm=cmr8

\def\qed{\hbox{${\vcenter{\vbox{			
   \hrule height 0.4pt\hbox{\vrule width 0.4pt height 6pt
   \kern5pt\vrule width 0.4pt}\hrule height 0.4pt}}}$}}

\begin{document}

\title {
\begin{flushright}
{\normalsize IIT-HEP-96/4\\
\vspace {-.15 in}
hep-ex/9610003
}
\end{flushright}
\vskip 0.2in
\normalsize\bf CHARMONIUM PRODUCTION IN FERMILAB E789\footnotemark }

\author {\normalsize DANIEL M. KAPLAN             
      \\ {\small\it Illinois Institute of Technology} 
                   \\  \\ 
{\normalsize for the E789 collaboration\footnotemark}
         }  
%
\date{}

\maketitle

\begin{abstract}

Using a sample of $>10^5$ 
$J/\psi\to\mu^+\mu^-$ decays,
Fermilab experiment 789 has studied production of $J/\psi$ and $\psi^\prime$
in 800\,GeV proton-nucleon collisions. 
Differential cross sections and nuclear dependences
have been measured for charmonium as well as for
charm and beauty production.
While charm and beauty production are consistent with perturbative QCD 
calculations,
charmonium cross sections exceed the predictions of the color-singlet model by 
large factors, suggesting that additional mechanisms (such as color-octet 
production) may play important roles.
Nuclear dependences of production cross sections may offer a new tool for the 
detailed understanding of charmonium production.

\end{abstract}

\tolerance=20000

\section{Introduction}

\setcounter{footnote}{1}
\footnotetext{
Invited talk presented at the Quarkonium Physics Workshop, University of 
Illinois at Chicago, June 13--15, 1996.}
\addtocounter{footnote}{1}
\footnotetext{
E789 collaboration: L.~D.~Isenhower, M.~E.~Sadler,
R.~Schnathorst (Abilene Christian University);
Y.~C.~Chen, G.~C.~Kiang, P.~K.~Teng
(Academia Sinica, Taiwan);
G.~Gidal, P.~M.~Ho, M.~S.~Kowitt, K.~B.~Luk, D.~Pripstein
(University of California at Berkeley/LBNL);
L.~M.~Lederman,\footnotemark~ M.~H.~Schub
(University of Chicago);
C.~N.~Brown, W.~E.~Cooper, K.~N.~Gounder, C.~S.~Mishra
(Fermilab);
T.~A.~Carey, D.~M.~Jansen, R.~G.~Jeppesen, J.~S.~Kapustinsky, D.~W.~Lane, 
M.~J.~Leitch, P.~L.~McGaughey, J.~M.~Moss, J.~C.~Peng 
(Los Alamos National Laboratory);
D.~M.~Kaplan,\addtocounter{footnote}{-1}\footnotemark~
W.~R.~Luebke,\addtocounter{footnote}{-1}\footnotemark~
  R.~S.~Preston, J.~Sa, V.~Tanikella
(Northern Illinois University);
R.~Childers, C.~W.~Darden, J.~R.~Wilson
(University of South Carolina).
}
\footnotetext{Present address: Illinois Institute of Technology.}

Fermilab experiment 789 is a study of two-prong decays of beauty and charm 
which took data during the 1990/1 fixed-target run.
E789 has published 
the most precise measurement of the charm production cross section 
and its $A$ dependence in 800\,GeV proton collisions,\cite{Leitch} as well as
novel measurements at very forward $x_F$
of the $J/\psi$ production cross section 
and its $A$ dependence using our beam dump (or a beryllium insert
just upstream of the beam dump) as the target.\cite{Kowitt}
More recent results include
the only measurement to date of the cross section for beauty production in
proton collisions at fixed-target energy,\cite{Jansen}
observed via the process $b\to J/\psi+X,~J/\psi\to\mu^+\mu^-$,
and high-statistics studies of
$J/\psi$ and $\psi^\prime$ production.\cite{Schub,Leitch2}

\section{Apparatus Description}

\begin{figure}[htb]
\vspace{-1.4 in}
\centerline{\hspace{0.318 in}\epsfysize = 2.87 in \epsffile {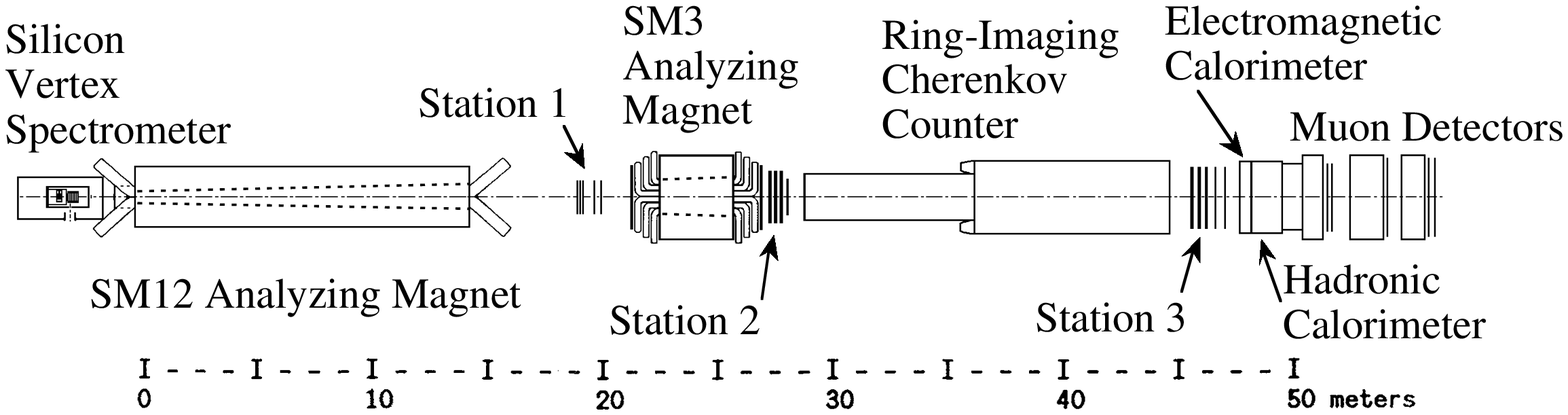}}
\vspace{-0.25 in}
\caption {Plan ($x$-$z$) view of E789 apparatus.\label{fig:E789}}
\vspace{-0.1 in}
\end{figure}

\begin{figure}[htb]
\vspace{-.5in}
\centerline{\hspace{0.4 in}\epsfysize = 2.54 in \epsffile {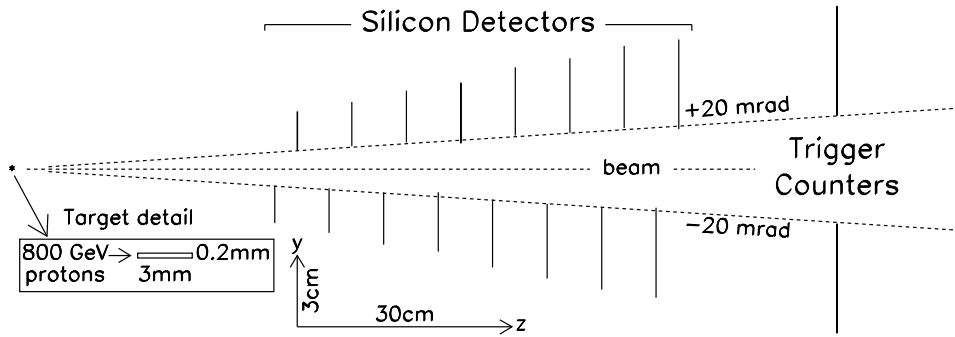}}
\vspace{-.05in}
\caption {Elevation view of E789 vertex telescopes; note that
the target dimensions indicated correspond to the beauty running,
and thinner targets were used for the charm running.\label{fig:ssd789}}
\end{figure}

The E789 apparatus (shown schematically in Fig.~\ref{fig:E789}) has been 
described in detail elsewhere\cite{Schub,E605}; we summarize briefly
here. It is based on the
pre-existing E605/772 spectrometer,\cite{E605} upgraded for this run by the
addition of two silicon-microstrip vertex telescopes (Fig.~\ref{fig:ssd789}), 
one above and one below the beam,
replacement of the Station-1 MWPCs with drift chambers,
and a tenfold increase in data-recording capacity (to 50\,MB/spill).
Wire-like targets were used to localize the primary-interaction vertex
in $y$ and $z$,
so that only the decay vertex need be reconstructed.

The spectrometer features two large analysis magnets, SM12 and SM3, which 
deflect charged particles in opposite directions in $y$. 
A water-cooled copper beam dump suspended within SM12 absorbs non-interacting
beam protons as well as secondaries emitted within $\approx$$\pm20\,$mr of the
beam in $y$. Shielding within and around SM12 absorbs neutral secondaries.
This geometry limits the pair
acceptance to $\approx$1\% but allows operation at 
high interaction rates.
The vertex telescopes and the
23 planes of scintillation-counter hodoscopes and drift chambers at
Stations 1, 2, and 3 measure the tracks of charged particles passing above 
or below the beam dump. The SM3 magnet 
serves to remeasure charged-particle momenta and thus to confirm the
target origin of tracks, allowing the copious background of muons created
within the beam dump to be rejected.
Particles are identified by electromagnetic and hadronic calorimeters, 
scintillation-hodoscope and proportional-tube muon
detectors, and a ring-imaging Cherenkov counter. 
In most events only two oppositely-charged particle tracks traversing the
spectrometer are fully
reconstructed, one passing through each vertex telescope.

Data were taken separately in charm and beauty spectrometer settings.
In the charm running we were able to operate the spectrometer at interaction 
rates up to 5\,MHz, using an on-line vertex trigger processor\cite{789Proc} 
to reject $>$80\% of hadron pairs from the target. The higher 
SM12 current used in the beauty setting allowed
$J/\psi$ and $\psi^\prime$ data to be taken at a 50\,MHz interaction rate, with 
no on-line vertex cut needed.

\section{Charm and Charmonium Production Cross Sections}
\label{E789-D}

\begin{figure}[htb]
\vspace{0.25in}
\centerline{\hspace{0.4in}\epsfysize = 2.4 in \epsffile {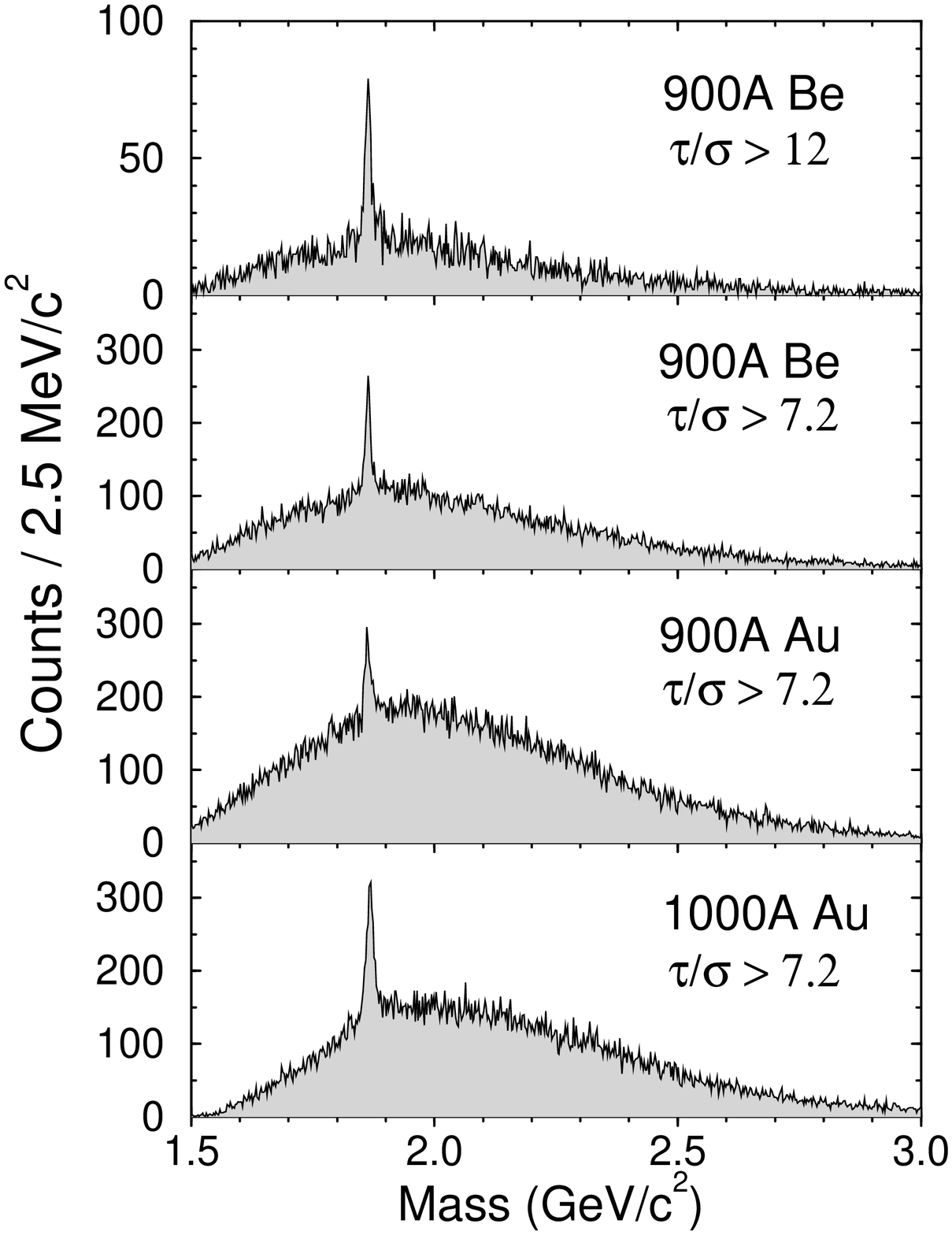}
\hspace{0.25in}\epsfysize = 2.65 in \epsffile {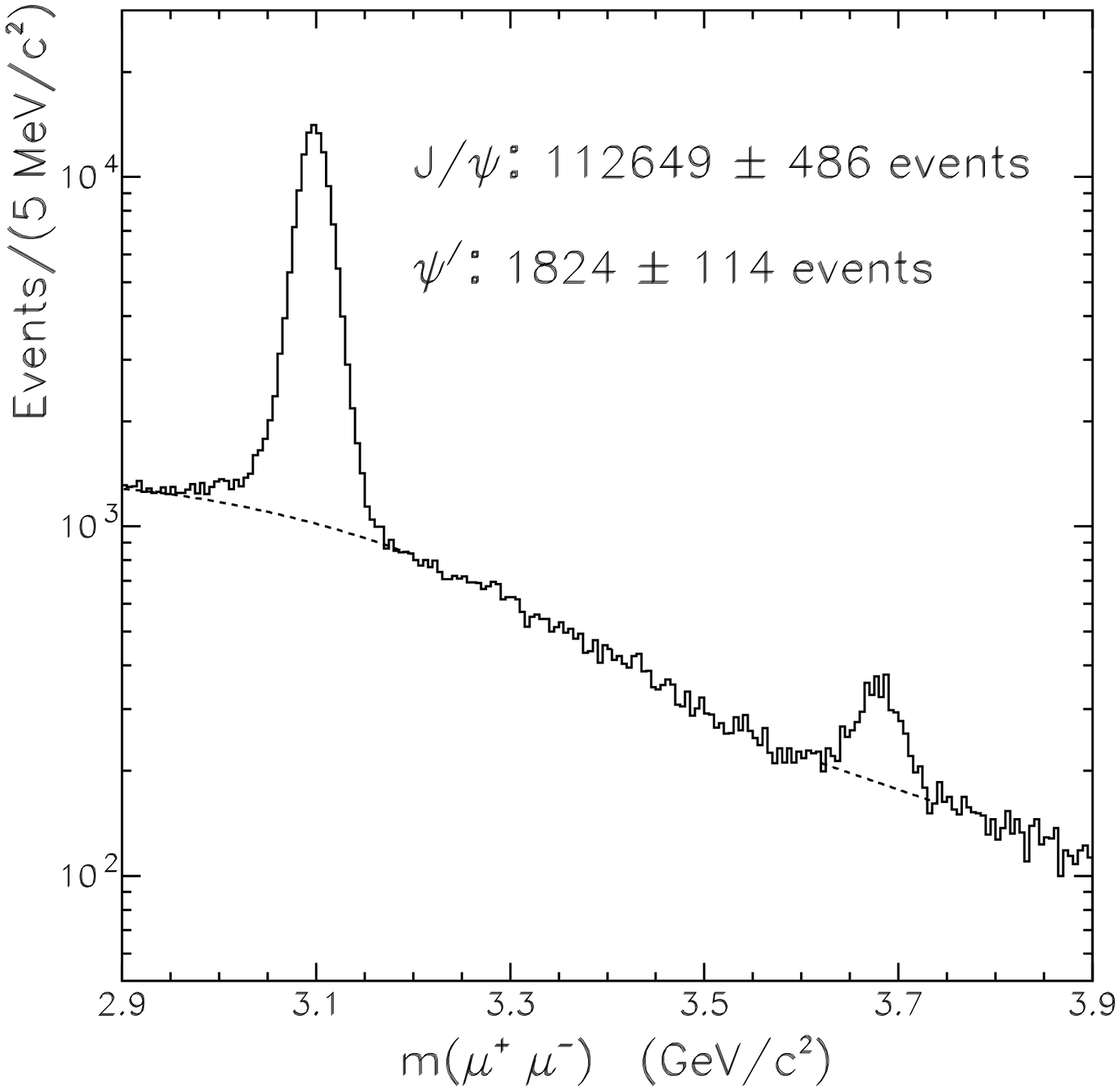}}
\vspace{0.25in}
\caption {Left: dihadron mass spectra for the E789 charm data samples for two 
different decay-vertex cuts.
Right: dimuon mass spectrum for E789 beauty data sample.\label{fig:mumu-mass}
\label{fig:D-peak}}
\end{figure}

We have measured differential cross sections for charm\cite{Leitch} and
charmonium\cite{Schub} production.
Charm data were taken using gold and beryllium targets at SM12 currents 
of 900 and 1000 amperes.
Fig.~\ref{fig:D-peak} shows the observed hadron-pair mass distributions (under
the $K^-\pi^+$ and $\pi^- K^+$ assumptions\footnote{The RICH detector was not
optimized for the $D$ mass region and was not used in this analysis.}~) 
for the
various E789 charm data samples, using two different decay-vertex cuts.
Clear $D^0 (\overline{D}{}^0)$ signals stand out above the dihadron background.
The looser vertex cuts ($\tau/\sigma_\tau>7.2$) were found to optimize 
the statistical significance of the
signal and were used in the cross-section and $A$-dependence analyses.

Our narrow range of 
acceptance in the longitudinal momentum of the pair (due to the 
beam-dump and shielding geometry) precludes a 
direct measurement of total cross section.
We measure\cite{Leitch} the $D^0/\overline{D}{}^0$ differential cross section 
$d \sigma/dx_F = 58\pm3\pm7\,\mu$b/nucleon at $\langle x_F \rangle = 0.03$.
This can be extrapolated over all $x_F$ (using the $x_F$ shape measured by 
previous experiments\cite{Ammar_Kodama}) to give
a total $D^0/\overline{D}{}^0$ cross section $\sigma = 
17.7\pm0.9\pm3.4\,\mu$b/nucleon.\cite{Leitch}
Averaging with previous measurements using 800\,GeV proton
beams\cite{Ammar_Kodama} 
gives
$\sigma(pN\to D^0\,X)
+ \sigma(pN\to {\overline D {}^0} \,X) = (20.9\pm 3.5)\,\mu$b/nucleon,
consistent with next-to-leading-order (NLO) QCD predictions\cite{Frixione}
 within the 
broad range of theoretical uncertainty.

%

\begin{figure}[htb]
\centerline{\epsfysize = 3.75 in \epsffile {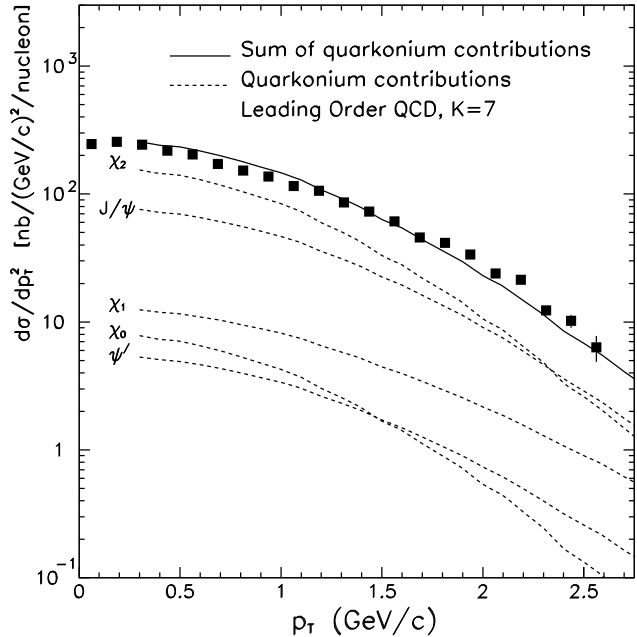}}
\caption {Differential cross section per nucleon
\vs $p_t$ for $J/\psi$ production 
compared with QCD prediction (note large phenomenological ``$K$" factor by
which the prediction has been multiplied). Dashed curves indicate 
indirect (via decay from higher-mass charmonium states) and direct 
contributions, and solid curve their sum.}
\label{fig:psi-pt}
\end{figure}

\begin{figure}[htb]
\centerline{\epsfysize = 3.75 in \epsffile {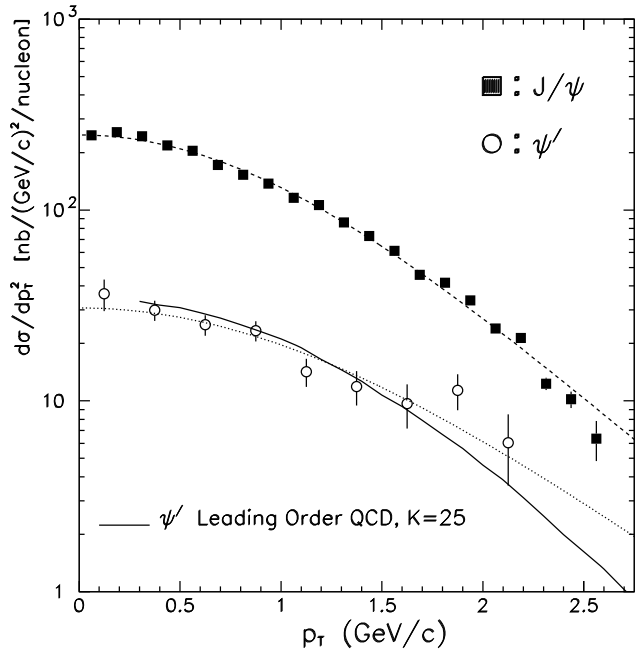}}
\caption {Differential cross section per nucleon
\vs $p_t$ for $\psi^\prime$ production 
compared with QCD prediction (note large phenomenological ``$K$" factor by
which the prediction has been multiplied). Also shown are phenomenological
fits to the $J/\psi$ and $\psi^\prime$ $p_t$ shapes of the form
$(1+(p_t/p_0)^2)^{-6}$; we find $p_0=3.00\pm0.02\,$GeV/$c$ for $J/\psi$
production and $3.60\pm0.32\,$GeV/$c$ for $\psi^\prime$.
\label{fig:psip-pt}}
\end{figure}

\begin{figure}[htb]
\centerline{\epsfysize = 3.75 in \epsffile {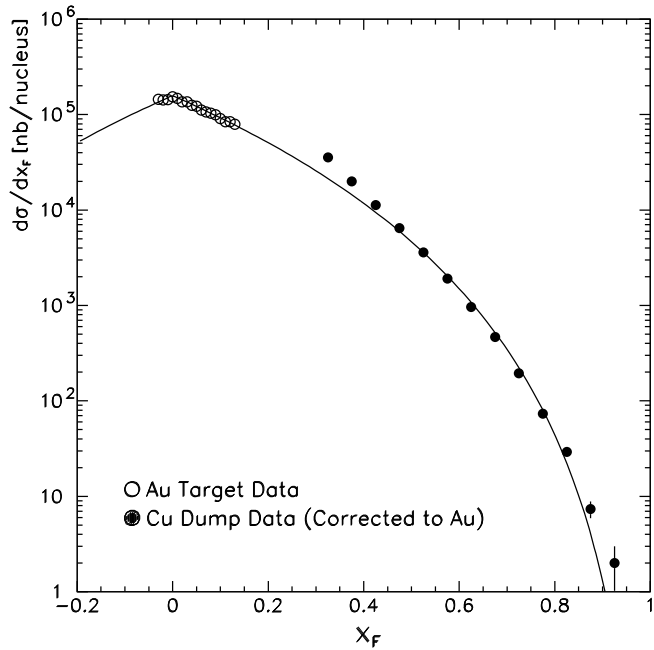}}
\caption {Differential cross section per nucleus 
\vs $x_F$ for $J/\psi$ production 
in 800\,GeV $p$-Au collisions along with phenomenological fit to the form
$(1-|x_F|)^n$; we find $n=5.0\pm 0.2$. (Note that lowest-$x_F$ points in 
beam-dump sample are suspect due to large 
corrections.\protect\cite{Kowitt-thesis})}
\label{fig:psi-xF}
\end{figure}

\begin{figure}[htb]
\centerline{\epsfysize = 3.75 in \epsffile {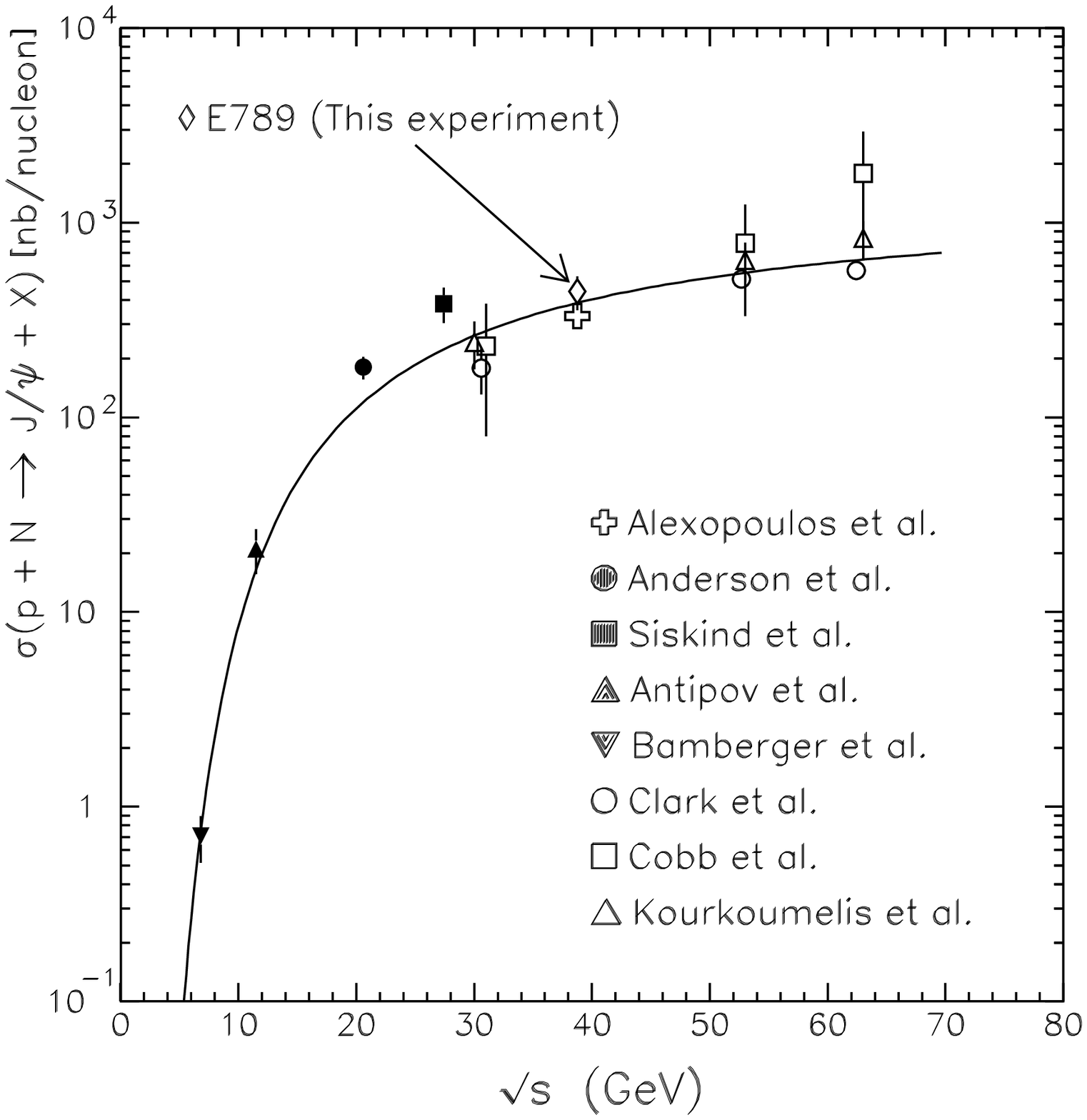}}
\caption {Energy dependence of total $J/\psi$ production 
cross section per nucleon; data from E789 and 
Refs.~\protect\cite{Alexopoulos}$^-$\protect\cite{Siskind}.}
\label{fig:sdep}
\end{figure}

Fig.~\ref{fig:mumu-mass} also shows the dimuon mass distribution from the
beauty running, and Figs.~\ref{fig:psi-pt} and \ref{fig:psip-pt} compare our
measured $J/\psi$ and $\psi^\prime$ differential cross sections with QCD
predictions.\cite{Psi-QCD}
Here there is little uncertainty in the extrapolation over $x_F$, with the 
$x_F$ shape well determined both by these ``open-aperture" data alone
and in combination with our beam-dump data,\cite{Kowitt} as indicated in 
Fig.~\ref{fig:psi-xF}.
As at the Tevatron Collider,\cite{Braaten}
$J/\psi$ and $\psi^\prime$ production are substantially underestimated
in the QCD calculation, with phenomenological ``$K$ factors" of 7 and 25 
(respectively) required to give agreement in magnitude between data and theory.
Note that the model calculation includes only contributions from
color-singlet charmonium states and neglects possible contributions from 
color-octet charmonium components\cite{Braaten,Mangano-ppbar}
 and from postulated states above $D\overline
D$ threshold.\cite{Page}
(These discrepancies have been the subject of much attention and are 
discussed further in Sec.~5.)
We find\cite{Schub} $\sigma(p+N\to J/\psi+X) = 442\pm2\pm88\,$nb/nucleon and
$\sigma(p+N\to \psi^\prime+X) = 75\pm5\pm22\,$nb/nucleon.
Comparison with previous results\cite{Alexopoulos}$^-$\cite{Siskind}
(Fig.~\ref{fig:sdep}) shows that the $J/\psi$ total cross section and its 
excitation curve are by now
well determined experimentally.

\section{Charm and Charmonium $A$ Dependence}

Nuclear dependences of production cross sections can shed light on production 
mechanisms and thus are of intrinsic interest. 
In addition, suppression of charmonium production
in nucleus-nucleus collisions has
been proposed as a signature for quark-gluon-plasma formation,\cite{Matsui} so
it is
important to study processes responsible for charmonium suppression in
proton-nucleus collisions, which might present a background to a
quark-gluon-plasma signal.
The production of heavy quarks is naively expected to depend
linearly on the atomic weight ($A$) of the target nucleus, since the dominant
QCD mechanisms (gluon-gluon fusion and $q\bar q$ annihilation)
involve hard partons.\cite{theory1,Frixione} However, in Fermilab
E772 we showed\cite{Alde}
that charmonium production in proton-nucleus collisions
in fact has a complicated dependence on $A$, 
parametrized as $d\sigma/dx_F\propto A^{\alpha(x_F)}$,
suggesting that other processes are at work in addition to those of 
perturbative QCD.

\begin{figure}[tb]
\vspace{0.25in}
\centerline{\hspace{-0.15in}\epsfysize=3.3in\epsffile {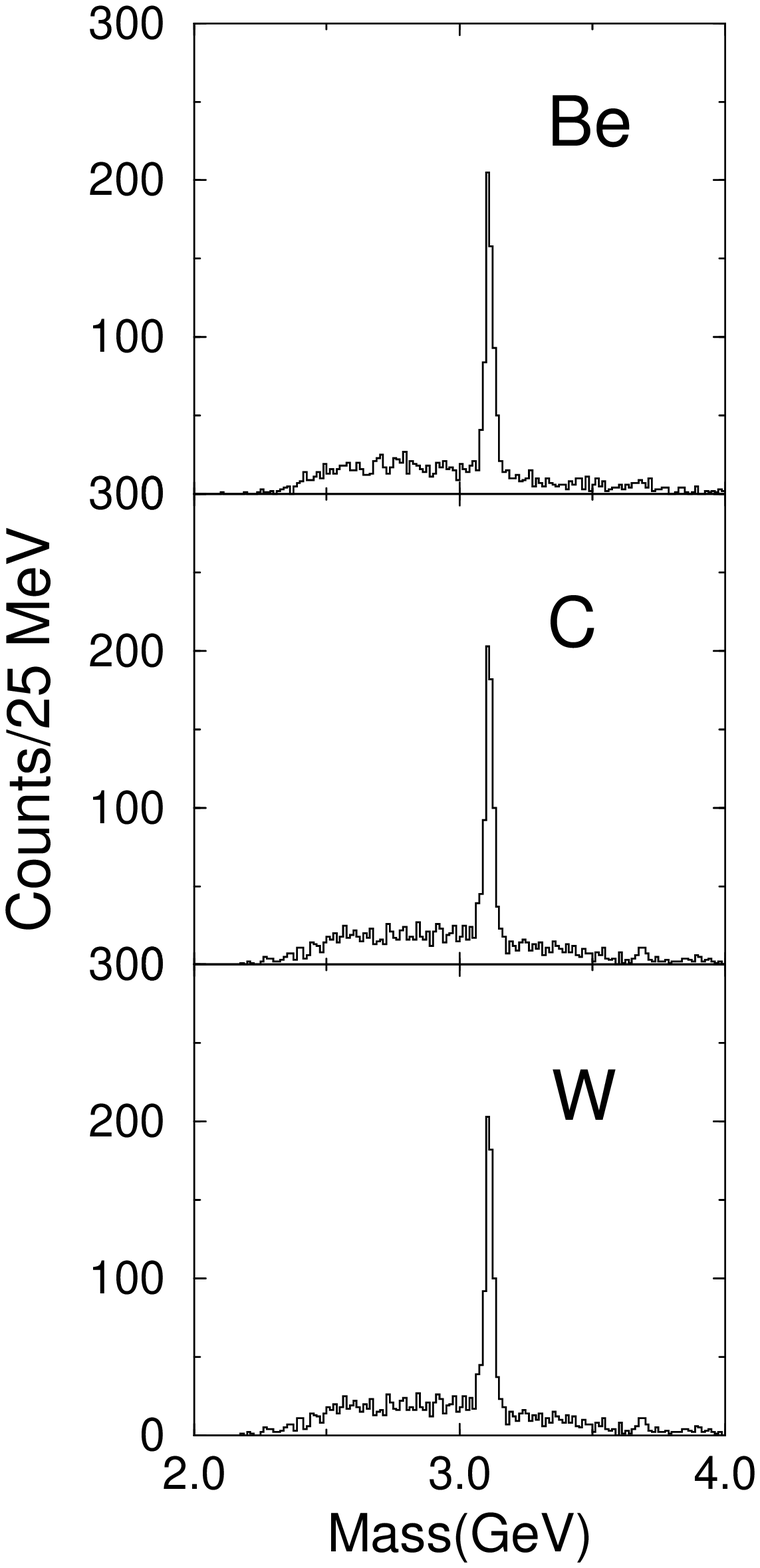}
\hspace{-0.1in}\epsfysize=3.3in\epsffile {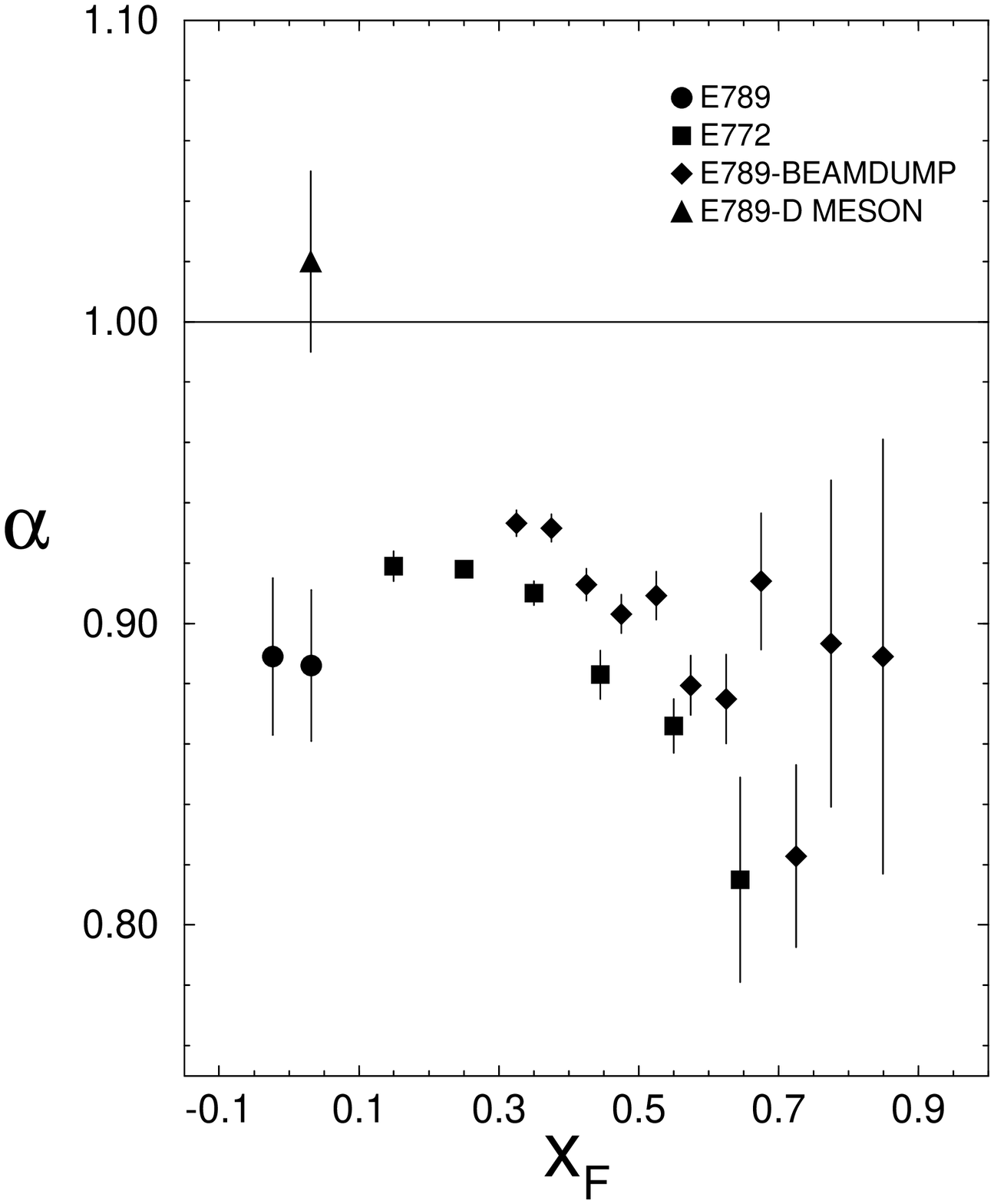}}
\caption {Left: dimuon mass spectra for E789 $J/\psi$ $A$-dependence data
samples.
Right: The exponent $\alpha$ of the $A$ dependence of the production cross
sections for $D^0(\overline{D}{}^0)$
and $J/\psi$ production \vs Feynman-$x$ as measured in E789; also shown are
$J/\psi$ results from E772.\label{fig:Adep}\label{fig:negjpsi}}
\end{figure}

The concentration of our charm sample 
in a narrow range of Feynman-$x$ results in the most precise 
determination to date of the charm-production nuclear dependence at a point in 
$x_F$, allowing a precise comparison to be made between open-charm and 
charmonium production. To augment the forward-$x_F$ $A$-dependence
measurements of E772, we took additional $J/\psi$ data in E789,
using a rotating wheel of beryllium, carbon, and tungsten targets placed 
1.27\,m downstream of the usual target position in order increase the 
acceptance near $x_F=0$. Fig.~\ref{fig:negjpsi} shows the dimuon mass
distributions thus obtained and the resulting $A$-dependence
exponent $\alpha$ \vs $x_F$ for $J/\psi$ compared with that for $D^0$.
 We see that (at least at small $x_F$) $\alpha$ is significantly lower for
charmonium than for charm: $D^0$ production depends linearly on $A$
($\alpha=1.02\pm0.03\pm0.02$ at $\langle x_F \rangle =0.03$),\cite{Leitch}
while for the $J/\psi$, $\alpha(0.03)=0.89\pm0.02$.\cite{Kowitt,Leitch2}

%
The increased nuclear suppression at low $x_F$ of charmonium as compared to
charm is consistent with models in which charmonium
production is suppressed in nuclei due to dissociation by interaction with
co-moving partons.\cite{Leitch2} Models in which the nuclear suppression of
charmonium production is an initial-state effect (\eg due to possible
shadowing or nuclear modification
of the gluon structure function) are disfavored, since they would
predict similar nuclear dependences for $J/\psi$ and $D$ production.

Effects on charmonium production at large $x_F$ due to intrinsic
charm\cite{Vogt-Brodsky} (the presence of virtual $c\bar c$ pairs in the
nucleon sea) and initial-state parton energy loss\cite{Gavin-Milana} have
also been postulated.
The predictions of the intrinsic-charm model have not been borne out by our
data since they feature significantly larger (and more strongly $A$-dependent)
cross-section contributions at the largest $x_F$ than are
seen.\cite{Kowitt}
The qualitative trend we observe at large $x_F$
(nuclear suppression increasing with $x_F$) is successfully
accomodated in models which take account of parton energy 
loss\cite{Gavin-Milana} in traversing nuclear matter. Since gluons should 
interact more strongly with matter than quarks, the parton energy-loss model
makes the intriguing prediction that at the highest Feynman-$x$ (where $q\bar
q$ annihilation dominates over gluon-gluon fusion), the nuclear suppression
should decrease.\cite{Jain-Ralston}
The apparent increase of $\alpha$ at $x_F\approx0.8$ suggests that 
this may be occurring, but better statistics are needed for confirmation; these
should be forthcoming from Fermilab E866, which is to take data during the 
1996/7 fixed-target run.

\section{Discussion}
\label{sec:quarkonium}

Production of charm\cite{Leitch} and beauty\cite{Jansen}
quarks is in reasonable agreement with perturbative QCD calculations. On the
other hand, production of charm and beauty {\em quarkonia} are observed at
rates from one to two orders of magnitude higher than naively predicted.
The experimental facts of enhanced quarkonium
production in fixed-target experiments are not new, dating back to the 
mid-1970s when $J/\psi$ hadroproduction was first observed.
At that time it was realized that lowest-order production of $c\bar c$ in a
color-singlet state (via one intermediate virtual photon or three virtual
gluons) had difficulty accounting for the large cross sections
measured,\cite{Baier-Ruckl}
and the ``color evaporation"\cite{local-duality}  (or local-duality)
mechanism achieved currency: the $c\bar c$ pair could be produced in a colored
state and later emit a soft gluon to neutralize its color at little or no cost
in probability.

Recent advances in perturbative QCD have made the predictions of the 
color-singlet model computable with no free 
parameters,\cite{Mangano-Glasgow} allowing it to be definitively
ruled out by both fixed-target and collider\cite{CDF-Jpsi,D0-Jpsi} 
measurements. In the regime accessible to fixed-target experiments,
questions may still remain as to the applicability of factorization and the
role of intrinsic parton $k_t$. But the collider data
in the previously-inaccessible regime of $p_t\gg m$
(where perturbative calculations ought to be most trustworthy) 
have forced theorists to consider seriously additional non-perturbative 
mechanisms.
As a result, the leading
candidate models which have emerged are the color-octet 
model\cite{Braaten,Mangano-ppbar} 
and an updated
color-evaporation model.\cite{evaporation}
 These models, while less predictive than 
the color-singlet model, nevertheless make strong predictions (for example, 
that $J/\psi$'s should be highly polarized at high $p_t$), which can be 
tested in detail in upcoming experiments. In this connection we mention
Fermilab E866, which should record $>10^6$ $J/\psi\to\mu^+\mu^-$ decays (in
closed aperture) in the upcoming run, as well as the C0 Charm
project,\cite{C0-charm} in which a sample of $\sim10^7$ $J/\psi$ decays could
be accumulated, with (due to the open geometry)
most final-state particles accompanying the $J/\psi$ also measured.
Thus C0 Charm holds the possibility of high-statistics measurements of 
$\chi_c$ (as well as $J/\psi$, $\psi^\prime$, and open-charm) production at
fixed-target energy.

Further data on $A$ dependences could also be useful.
Energy loss in nuclear matter may be a means to distinguish color-singlet and 
color-octet charmonium states, since the color-octet state has gluonic
quantum numbers and may be strongly absorbed.\cite{Heinz-plm} 
Distinguishing initial- and 
final-state $A$-dependence mechanisms calls for more data on the $A$ 
dependence of open-charm production at large $x_F$, another area in which 
C0 Charm could contribute.

\end{document}